\input{aipcheck}

\documentclass[
    ,final            
  ]
  {aipproc}

\layoutstyle{6x9}

\begin{document}

\title{A Search for Short Time-Scale Optical 
Variability in the GRB 030329 Afterglow}

\author{N.\ Mirabal}{
  address={Astronomy Department, Columbia University, 550 West 120th Street,
 New York, NY 10027}
}

\author{J. P.\ Halpern}{
  address={Astronomy Department, Columbia University, 550 West 120th Street,
 New York, NY 10027}
}

\author{M. Bureau}{
  address={Astronomy Department, Columbia University, 550 West 120th Street,
 New York, NY 10027}
}

\author{K. Fathi}{
  address={Kapteyn Astronomical Institute, 
Postbus 800, 9700 Av, Groningen, The Netherlands}
}

\begin{abstract}
We present a densely sampled $R$-band 
light curve of the optical afterglow of GRB
030329 obtained during the period 15.5--23 hours after the 
burst. This dataset allows us to search for short time-scale 
fluctuations that 
must be present at some level due to inhomogeneities in the 
circumburst medium and/or if the afterglow is intrinsically variable.
\end{abstract}

\date{\today}

\maketitle


\section{Introduction}
The recent detection of early fluctuations about the mean power-law 
decay of GRB afterglows \cite{Bersier:2003, Mirabal:2003}
may provide important clues in our understanding of the 
afterglow evolution. However, since several interpretations have been 
put forth to explain the fluctuations (e.g. \cite{Heyl:2003}), it 
is crucial to determine the mechanism 
ultimately responsible for the variability. Here we present an 
initial attempt to help constrain the microphysics of GRB afterglows
using a rapid optical photometric sequence. 

\section{Discussion}

High-speed optical photometry of 
the GRB 030329 afterglow  
was obtained at the MDM 1.3 m telescope beginning on March 30 03:05 UT,
15.5 hours after the burst.  
The data consists of a 7.5--hour time series in the $R$-band with
a time resolution of 90 s. Altogether, we obtained 
a total of 306 points. During this period, the magnitude of the
optical afterglow declined from
$R = 15.4$ to $R = 16.2$. Figure 1 shows 
a power-law decay slope $\alpha$ = --1.931 $\pm$ 0.005 fitted to the data, where the
uncertainties are dominated by faint comparison stars and large
airmasses at the beginning and the end of the run. 

The computed residuals 
indicate that the average fluctuation about the power-law fit  
is smaller than 0.01 mag on time scales of minutes to 
hours. If the flux variability $f_{1}/f_{0}$ 
traces the density contrast $n_{1}/n_{0}$ as a function of 
$f_{1}/f_{0} \propto (n_{1}/n_{0})^{1/2}$ \cite{Lazzati:2002}, this implies
that the enhancements in the vicinity of the GRB 
are less  than a factor of 1.03 in density.  In other words, the circumburst 
medium was close to homogeneous between 1.4 $\times 10^{17}$ cm and 1.7 
$\times 10^{17}$ cm from the burst site, 
assuming a spherical adiabatic blast wave expanding 
in a stellar wind. 
Unfortunately this fast sequence by itself cannot place
strong limits on intrinsic afterglow variability, but it shows that 
the afterglow evolution can be highly uniform on short time-scales.

\begin{figure}
\includegraphics[height=.6\textheight]{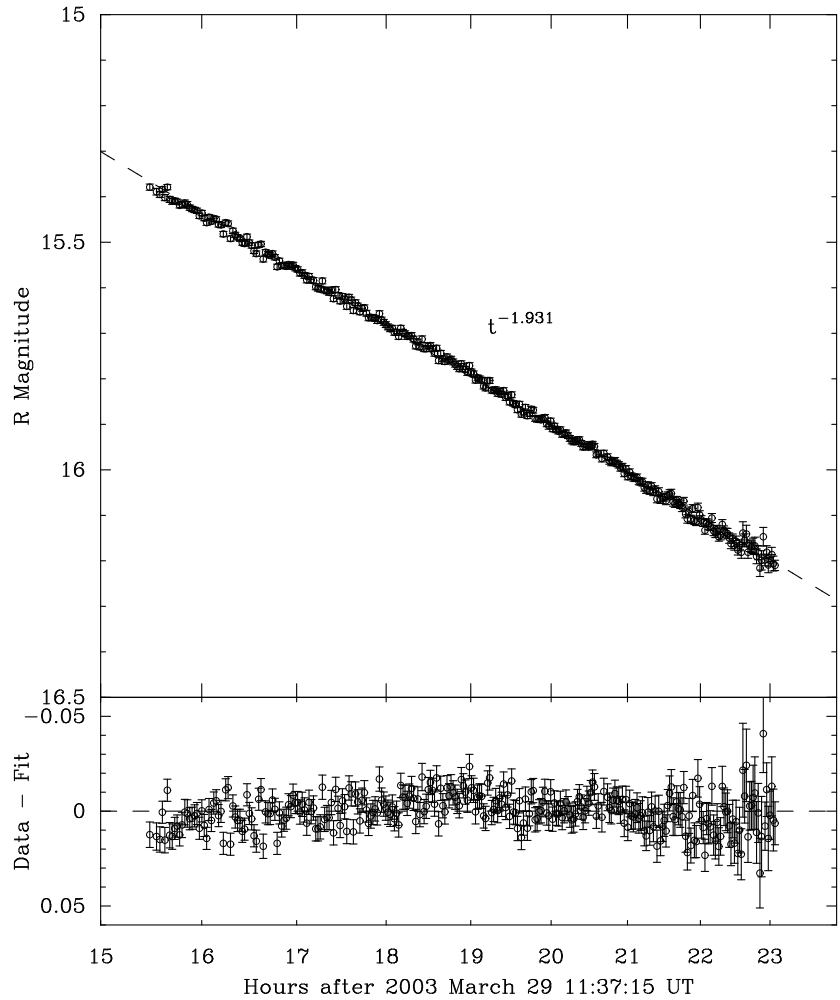}
  \caption{{\it Top}:
$R$-band light curve of the GRB 030329 afterglow. 
{\it Bottom}: Difference between data and model.
}
\end{figure}

\section{Conclusions and Future Work}

We have shown preliminary results of a high-speed optical photometry program
to access short time-scale fluctuations in 
GRB afterglows. A comprehensive effort to observe and model fluctuations 
at different time scales might hold the key to understanding the 
microphysics of GRB afterglows.


\IfFileExists{\jobname.bbl}{}
 {\typeout{}
  \typeout{******************************************}
  \typeout{** Please run "bibtex \jobname" to optain}
  \typeout{** the bibliography and then re-run LaTeX}
  \typeout{** twice to fix the references!}
  \typeout{******************************************}
  \typeout{}
 }

\end{document}